# Deriving Time-varying Cellular Motility Parameters via Wavelet Analysis

Yanping Liu,[1,*] Yang Jiao,[2,*] Guoqiang Li,[1] Gao Wang,[1] Jingru Yao,[1] Guo Chen,[1] Silong Lou,[3] Jianwei Shuai,[4,†]  Liyu Liu[1,†]

[1] Chongqing Key Laboratory of Soft Condensed Matter Physics and Smart Materials, College of Physics, Chongqing University, Chongqing, 401331, China
[2] Materials Science and Engineering, Arizona State University, Tempe, Arizona 85287, USA
[3] Department of Neurosurgery, Chongqing Cancer Hospital, Chongqing, 400030, China
[4] Department of Physics, Xiamen University, Xiamen, 361102, China

[*]These authors contributed equally to this work.
[†]Corresponding authors: jianweishuai@xmu.edu.cn, lyliu@cqu.edu.cn

**ABSTRACT**

Cell migration is an indispensable physiological and pathological process for normal tissue development and cancer metastasis, which is greatly regulated by intracellular signal pathways and extracellular microenvironment (ECM). However, there is a lack of adequate tools to analyze the time-varying cell migration characteristics because of the effects of some factors, i.e., the ECM including the time-dependent local stiffness due to microstructural remodeling by migrating cells. Here, we develop an approach to derive the time-dependent motility parameters from cellular trajectories, based on the time-varying persistent random walk model. In particular, we employ the wavelet denoising and wavelet transform to investigate cell migration velocities and obtain the wavelet power spectrum. The time-dependent motility parameters are subsequently derived via Lorentzian power spectrum. Our analysis shows that the combination of wavelet denoising, wavelet transform and Lorentzian power spectrum provides a powerful tool to derive accurately the time-dependent motility parameters, which reflects the time-varying microenvironment characteristics to some extent.

**Keywords**: cell migration, time-varying microenvironment, motility parameter, Lorentzian power spectrum, wavelet transform.

**I. INTRODUCTION**

Cell migration [1] is a ubiquitous and basic biological phenomenon that underlies many crucial physiological processes for normal tissue and organ development as well as immunological responses [2], wound healing [3], embryogenesis [4]. Eukaryotic cell migration is a complex process involving many cellular and sub-cellular level events [5], which are regulated by various cellular signaling pathways [6] and the external microenvironment [7]. The onset of ill-regulated cell migration is often associated with many human diseases, and the most representative example is cancer metastasis [8,9].

In order to study cell behaviors in microenvironment, a number of interesting works have been done in recently years. It was reported that the substrates with different rigidities lead to different cell movements, namely stiffer substrates generally promote the directionality of cell movement, while soft substrates typically result in random motions [10]. These distinct behaviors regulated by substrate stiffness are called as "durotaxis", which are combined with mechanical strains to control a number of pathological processes involving cell migration [11]. In addition to substrates, the gradient of the nanoscale topographic features in microenvironment will guide a new type of directed migration termed as "topotaxis" and the direction of topotaxis can reflect the effective cell stiffness [12]. Furthermore, in order to systematically investigate the heterogeneous microenvironment, a micro-fabricated biochip was constructed to create a 3D funnel-like matrigel interface, which verified that the heterogeneous structures of microenvironment can guide the aggressive cell invasion in the rigid matrigel space [13]. More recently, it was shown that the local fiber alignment in a constructed collagen I–matrigel microenvironment directs the migration of MDA-MB-231 breast cancer cells during the intravasation into rigid matrigel [14].

To phenomenologically describe the anisotropic migratory behaviors, a persistent random walk model (PRW) [15-17] has been proposed, which explicitly considers the memory of cell to the past velocities. The PRW model is based on Brownian motion [18], and can be derived from Langevin equation [19] of the following form

$$\frac{d\vec{v}}{dt} = -\frac{\vec{v}}{P} + \frac{S}{\sqrt{P}} \cdot \tilde{w}, \qquad (1)$$

where $\vec{v}$ is the migration velocity, P the persistence time, S the averaged migration speed and $\tilde{w}$ the random vector of a Wiener process [20]. Note that neither of the parameters P and S change with time in PRW model. Inspired by the PRW model, many novel models have been constructed for exploring how cells behave in complex microenvironment [21]. For example, amoeba exhibits a special random walk mode, which can greatly increase the chance of finding a target [22]. Likewise, CD8 (+) T cell in brain performs a movement known as generalized Lévy walk, which enables T cells to find rare targets [23]. Moreover, a mathematical model was developed for describing the statistical properties of cell's velocity and centroid, which are consistent with the phenomenological description of amoeboid motility [24].

In the study of cellular phenomena and modeling cell movement, accurately characterizing cell migration capability is of great interest. In order to address the challenge that the total time of the recorded trajectories in the experiments may not be precisely controlled, an optimal estimation was constructed to obtain the diffusion coefficients based on the individual and short trajectories [25]. Similarly, an unbiased and practically optimal covariance-based estimator was also constructed to optimally determine the diffusion coefficient of a diffusing particle from a time-lapse recorded trajectory [26]. Besides the diffusion coefficient, the direction autocorrelation function and other essential parameters are computed to analyze cell migration in two dimensions, based on an open-source computer program, DiPer [27]. In the previous works, we also developed exclusive methods to analyze anisotropic microenvironment and derived the time-independent cellular motility parameters, namely persistence time P and migration speed S [28,29].

Different from the cases above whose characteristics and properties are assumed to be non-varying with time, the more complex microenvironment changes globally or locally due to the changes of temperature, pressure, the heterogeneous surfaces in which cells migrate [10,11], the special components (oriented fibers) [14] or the concentration of biochemical factors [7] such as cytokine or drug molecules. In addition, migrating cells can actively remodel the microenvironment either mechanically or chemically, leading to spatial-temporally varying properties that in turn influences cell migration [30-32]. Accordingly, there are a few works focusing on time-varying cell dynamics in heterogeneous microenvironment. For instance, when cells migrate on the tissue with cultured polystyrene surface, a random motion coefficient increased significantly over time, while for experiments with untreated polystyrene plates, the random motion coefficient remained relatively constant [33]. Moreover, a superstatistical approach was proposed to derive the time-dependent statistical parameters (persistence and activity) from time-lapse recorded trajectories [34]. Although the approach enable one to derive the persistence and activity [cf. Eq. (24)], other important dynamical parameters such as the time-dependent persistence time P and migration speed S [cf. Eq.(1)] cannot be derived using this approach.

In this study, we consider the persistent random walk in microenvironment with time-varying characteristics, and propose an approach to derive the time-dependent motility parameters (persistence time and migration speed) from cell migration trajectories. Specifically, we introduce wavelet transform to analyze the cell migration velocities and obtain the wavelet power spectrum, which exhibits the time-frequency characteristics of cell trajectories. Moreover, the time-dependent motility parameters can be derived from the fits to wavelet power spectrum at each moment with Lorentzian power spectrum. Finally, we find that the wavelet denoising on migration velocities will contribute to derive more accurate motility parameters before performing the wavelet transform.

The rest of the paper is organized as follows: In Sec. II, we introduce the PRW model in microenvironment with time-varying characteristics, further explore the properties of cell migration and illustrate the limitations of commonly used physical quantities for characterizing cellular dynamics in this situation. In Sec. III, we combine the wavelet denoising, wavelet transform and Lorentzian power spectrum to derive the time-dependent motility parameters, and demonstrate the utility of and validate the proposed approach via representative examples of time-varying microenvironments. In Sec. IV, we analyze the effects of several factors on the performance of the approach, including the changing rates of microenvironmental properties (motility parameters), the number of the recorded cells ($N_c$), the total recording time for individual trajectories (T), and the sampling time interval ($\Delta T$), and clarify the limitations of our approach. In Sec. V, we provide concluding remarks.

## II. PERSISTENT RANDOM WALK MODEL WITH TIME-VARYING MOTILITY PARAMETERS

In this section, we develop a motility model to describe cell dynamics in microenvironment with time-varying characteristics. Based on the model, we computationally generate cell migration trajectories and explore the effect of microenvironment on cell behaviors. Based on the analysis of these trajectories, we show that commonly used classic physical quantities, including mean squared displacement (MSD), velocity autocovariance function (VC) and Lorentzian power spectrum (PS), are not sufficient to characterize cell motility.

### A. THE EFFECT OF MICROENVIRONMENT WITH TIME-VARYING CHARACTERISTICS ON CELL MOTILITY

Inspired by the time-varying physical and chemical properties in microenvironment, we generalize the classical persistent random walk model (PRW) [15-17] and obtain a new motility model, namely the time-varying persistent random walk model (TPRW). Note that both the parameters P and S in TPRW model are varying with time, which together quantify the time-dependent cell migration capability.

For simplicity, we first construct the following affine functions of time for P and S through Ref. [34], written as

$$P(t)=K_P \cdot t+P_0, \quad (2)$$

$$S(t)=K_S \cdot t+S_0, \quad (3)$$

where the constant $P_0$ and $S_0$ are motility parameters at t = 0. $K_P$ and $K_S$ are coefficients quantifying the changing rates of motility parameters P and S, both of which reflect the changes of microenvironmental properties to some extent. The functions defined above indicate that the cellular migration capability gradually increases with time, corresponding to the real-time enhancement of the microenvironment to cell migration.

### B. NUMERICAL SIMULATION OF CELL MIGRATION TRAJECTORIES BASED ON TPRW MODEL

In order to explore the characteristics of cell migration trajectories, we first specify the parameters in time-dependent functions in Eqs. (2-3), i.e., $K_P$ = 7.292e-4, $P_0$ = 0.3 min, $K_S$ = 2.083e-4 µm/min$^2$ and $S_0$ = 0.1 µm/min. Here, the values of parameters are defined by referring the works [34,35], partly. Thus, the motility parameter P lies in the interval of 0.3 ~ 1.0 min, while S in the interval of 0.1 ~ 0.3 μm/min, as plotted in Fig. 1(a). Cell trajectories can be then simulated by TPRW model according to Eqs. (4-9). In particular, the cell position at each time step can be obtained easily according to the following equations [29,36]

$$x(t+\Delta t)=x(t)+\Delta x(t,\Delta t), \quad (4)$$

$$y(t+\Delta t)=y(t)+\Delta y(t,\Delta t), \quad (5)$$

here $\Delta x$ and $\Delta y$ are displacements of cell position in x and y axes in the time step size of $\Delta t$. Further, the displacements are given by

$$\Delta x(t,\Delta t)=\alpha(t)\cdot\Delta x(t-\Delta t,\Delta t)+F(t)\cdot\tilde{W}, \quad (6)$$

$$\Delta y(t,\Delta t)=\alpha(t)\cdot\Delta y(t-\Delta t,\Delta t)+F(t)\cdot\tilde{W}, \quad (7)$$

where $\alpha(t)=1-\Delta t/P(t)$ and $F(t)=\sqrt{S(t)^2\cdot\Delta t^3/P(t)}$. The former denotes the memory of cell to the past velocities, while the latter is noise amplitude. $\tilde{W}\sim N(0,1)$ is Gaussian white noise. Note that the values of P are not less than that of $\Delta t$, ensuring that α is always greater than zero. In computer simulations, the total recording time T is 960 min and the time step size $\Delta t$ is 0.2 min. At the two limits of persistence time, the cell migration described by TPRW model either becomes the ballistic motion (P ~ infinity) or the random walk (P ~ $\Delta t$). In order to mimic the uncertainties in experimental observations, we further add the positioning errors $\sigma_{pos}$ to the simulated trajectories by

$$\hat{x}(t)=x(t)+\sigma_{pos}\cdot\tilde{W}, \quad (8)$$

$$\hat{y}(t)=y(t)+\sigma_{pos}\cdot\tilde{W}, \quad (9)$$

where $\sigma_{pos}$ is set as 0.01 μm [16]. We obtain 200 independent cell migration trajectories in numerical simulations, and a representative trajectory is presented in Fig. 1(b).

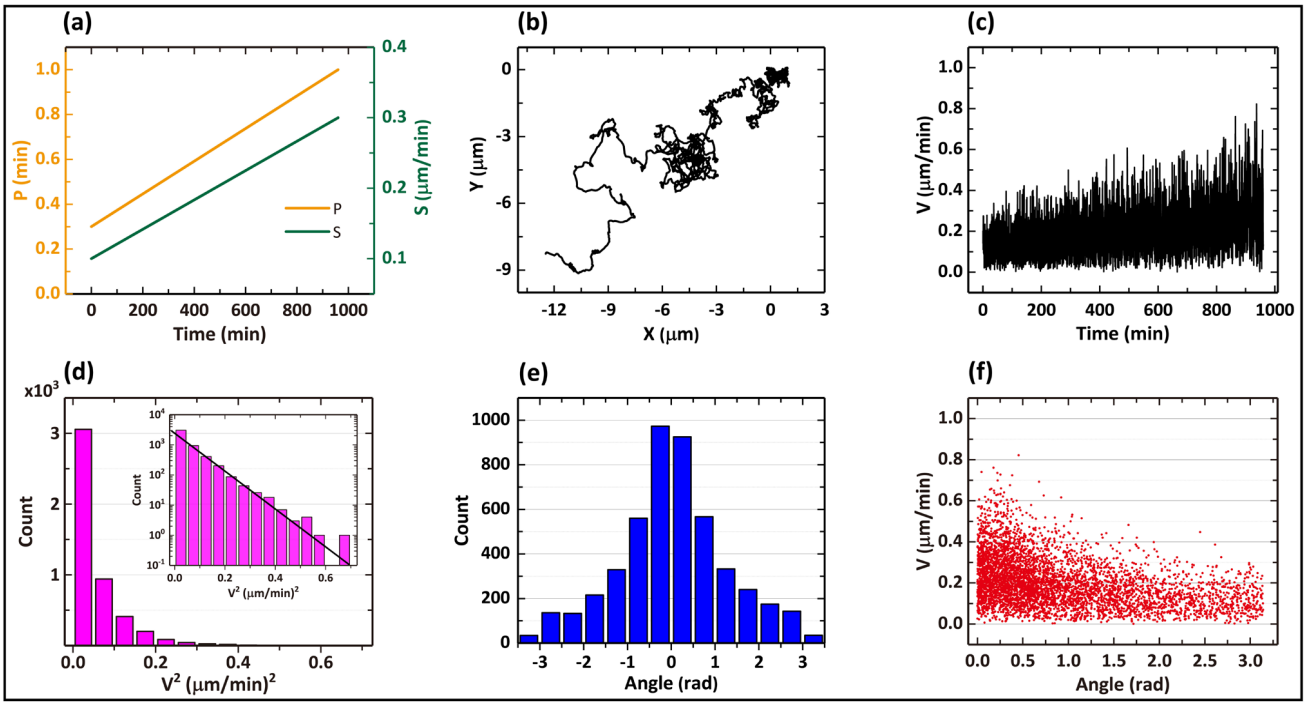

Fig. 1. The persistent random walk model with time-varying motility parameters (TPRW model). (a) The linearly time-dependent functions [cf. Eqs. (2-3)] of motility parameters. The orange line corresponds to the persistence time P, while the green line corresponds to the migration speed S. (b) Individual cell migration trajectories consisting of 4800 + 1 (N+1) frames simulated by TPRW model. (c) Cell migration velocities plotted against time. (d) Distribution of the square of cell migration velocities. The inset indicates the same histogram but in lin-log axes and the black straight line denotes an exponential decay. (e) Distribution of the angles between any two successive velocities. (f) Cell migration velocities as functions of angles between any two successive velocities.

For each trajectory, the migration velocities are computed based on the displacements within the time step $\Delta t$. The velocities in Fig. 1(c) gradually increase with time, which is the consequence of motility parameters in Fig. 1(a). Moreover, the square of velocities obeys an exponential decay in lin-log axes [see Fig. 1(d)], which indicates that velocity components still are Gaussian distributed when considering the positioning errors [35]. Based on the computed migration velocities, we also obtain the angle displacements between any two successive velocity vectors. The corresponding distribution is presented in Fig. 1(e), which are symmetric at 0 rad in the interval of (-π ~ π) and means that the chances of turning left and right are identical when cell migrates. Moreover, the distribution indicates that a self-propelled cell prefers to migrate along a fixed direction, instead of a large deflection. In addition, migration velocities become smaller when the corresponding angle displacements increase as depicted in Fig. 1(f). These results indicate that a high-speed moving cell tends to migrate in straight line, while a low-speed cell tends to make turn. The phenomenon is also investigated in Refs. [24,37]. These aspects indicate the TPRW model are consistent with the Ornstein-Uhlenbeck process (OU) [38] to some extent, which may be explained by the fact that the TPRW model can be regarded as the superposition of many OU models.

## C. CHARACTERIZING CELL MOTILITY DESCRIBED BY TPRW MODEL

In what follows, we employ three classical physical quantities, including the mean square displacement, velocity autocovariance function, and Lorentzian velocity power spectrum to investigate the overall averaged motility parameters P and S based on 200 cell migration trajectories. We illustrate the limitations of the three quantities when extracting the time-dependent motility parameters.

### 1. MEAN SQUARED DISPLACEMENT

First we calculate the mean squared displacement (MSD) [27,35,39] based on the coordinates of cell migration trajectory $\vec{r}_{i \cdot \Delta t}$ (i=0,...,N) in Cartesian coordinates given by

$$\text{MSD}(n \cdot \Delta t) = \frac{1}{N-n+1} \sum_{i=0}^{N-n} \left( \vec{r}_{(i+n) \cdot \Delta t} - \vec{r}_{i \cdot \Delta t} \right)^2, \quad (10)$$

where $\vec{r}$ is the position vector of individual cells at each time step, N the total number of displacements per trajectory, n the step size. The overall averaged MSD for 200 simulated trajectories are plotted in Fig. 2(a). Here the theoretical MSD [27,36] reads as

$$\text{MSD}(t) = 4D \cdot (t - P + P \cdot e^{-t/P}) + 4\sigma_{pos}^2, \quad (11)$$

where D is the diffusion coefficient. Therefore, we obtain a set of motility parameters P, D and $\sigma_{pos}$ by the fit to the overall averaged MSD, and the migration speed S is computed by the following formula [28,33]

$$S=\sqrt{\frac{2D}{P}}. \tag{12}$$

The resulting parameters are shown in Figs. 2(d-f).

## 2. VELOCITY AUTOCOVARIANCE

Similarly, Fig. 2(b) shows the overall averaged velocity autocovariance function (VC) for the 200 migration trajectories, in which the VC for individual cells is computed by [16]

$$VC(n\cdot\Delta t)=\langle \vec{v}_{i\cdot\Delta t}\cdot\vec{v}_{(i+n)\cdot\Delta t}\rangle=\frac{1}{N-n+1}\sum_{k=1}^{N-n}\left(\vec{v}_{k\cdot\Delta t}-\frac{1}{N-n}\sum_{l}^{N-n}\vec{v}_{l\cdot\Delta t}\right)\cdot\left(\vec{v}_{(k+n)\cdot\Delta t}-\frac{1}{N-n}\sum_{l=n+1}^{N}\vec{v}_{l\cdot\Delta t}\right). \tag{13}$$

After computing the VC based on cell migration velocities, a widely-used fit can be performed via a revised exponential decay, which is given as [35]

$$VC_j = VC_j^{(true)} \quad \text{for} \quad |j|\geq 2, \tag{14}$$

$$VC_{\pm 1} = VC_1^{(true)} - 2\sigma_{pos}^2/(\Delta t)^2, \tag{15}$$

$$VC_0 = VC_0^{(true)} + 4\sigma_{pos}^2/(\Delta t)^2. \tag{16}$$

Note that the theoretical VC only is affected at times $t_0=0$ and $t_{\pm 1}=\pm\Delta t$ when considering the positioning errors. Among the equations above, the $VC^{(true)}$ is defined by

$$VC_{j-k}^{(true)} = \frac{2P^2\cdot[\cosh(\Delta t/P)-1]}{(\Delta t)^2}\cdot VC(t_j-t_k) \quad \text{for} \quad j\neq k, \tag{17}$$

$$VC_0^{(true)} = \frac{2P^2\cdot(e^{-\Delta t/P}-1+\Delta t/P)}{(\Delta t)^2}\cdot VC(0) \quad \text{for} \quad j=k. \tag{18}$$

It is evident that the overall averaged VC in lin-log axes cannot be nicely fitted by the revised exponential decay [see Fig. 2(b)], indicating the insufficiency of the revised exponential decay in analysis of cell migration simulated by TPRW model. Even so, we can still obtain a set of parameters, as shown in Figs. 2(d-f).

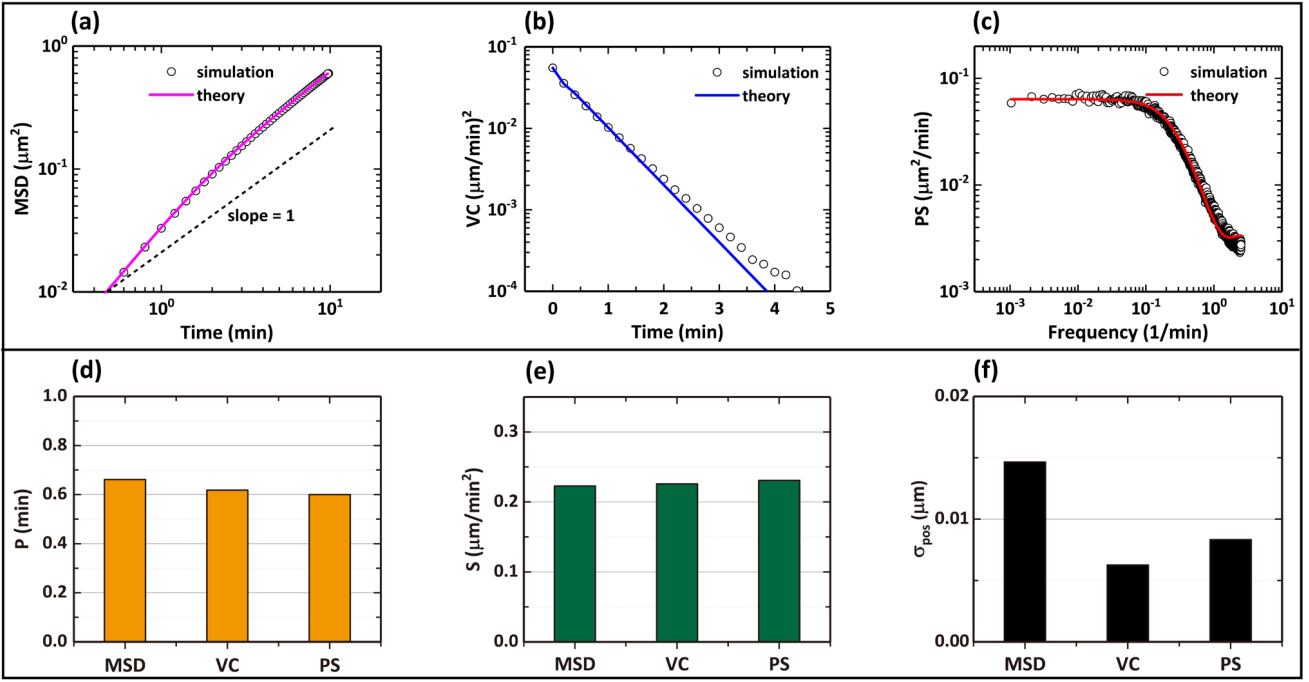

Fig. 2. The motility parameters derived from three overall averaged classical physical quantities. (a) Mean squared displacement of cell migration calculated with Eq. (10) and the mauve theoretical curve from Fürth's formula [cf. Eq. (11)]. The black dotted line is an auxiliary line with the slope 1. (b) Velocity autocovariance function computed from Eq. (13) and the blue theoretical curve from Eqs. (14-18). (c) Fourier power spectrum of migration velocities obtained from Ref. [35] and the red theoretical curve from Eqs. (19-21). (d) Comparison of the parameters P fitted from three physical quantities. (e) Comparison of the fitted parameters S. (f) Comparison of the fitted positioning errors $\sigma_{pos}$.

## 3. LORENTZIAN VELOCITY POWER SPECTRUM

It has been reported that the physical quantities used above could not return reliable errors on the fitted parameters P and S because of the correlations between the velocities [35]. Therefore, another method has been suggested to transform the time domain to frequency domain by performing Fourier transform (FT) of the VC according to Wiener-Khinchin theorem [40,41], which can eliminate the correlations. The results are referred as Fourier power spectrum (PS), as seen in Fig. 2(c). It behaves like that for PRW model in microenvironment with temporally non-varying properties. Then we can fit the Fourier power spectrum with the Lorentzian power spectrum of OU process [35], which is defined as

$$PS_u(f_k) = PS_u^{(true)}(f_k) + \frac{4\sigma_{pos}^2}{\Delta t}\left[1-\cos(\pi \cdot f_k/f_{Nyq})\right], \quad (19)$$

where the first term on the right side of Eq. (19) is the true expression of Lorentzian power spectrum, with the following form

$$PS_u^{(true)}(f_k) = \frac{(1-c^2)}{c} \cdot \left(\frac{P}{\Delta t}\right)^2 \cdot PS_v^{(aliased)}(f_k) + 4D \cdot \left(1 - \frac{1-c^2}{2c} \cdot \frac{P}{\Delta t}\right), \quad (20)$$

in which the term $PS_v^{(aliased)}(f_k)$ is defined by

$$PS_v^{(aliased)}(f_k) = \frac{\left\langle \left|\hat{\vec{v}}\right|^2 \right\rangle}{t_{msr}} = \frac{(1-c^2) \cdot 2D \cdot \Delta t / P}{1 + c^2 - 2c \cdot \cos(\pi \cdot f_k / f_{Nyq})}, \tag{21}$$

and the second term on the right side of Eq. (19) is an additional noise term when considering the effect of positioning noise. Here, $c = \exp(-\Delta t / P)$, $f_k = k \cdot \Delta f$ $(k=1,...,N/2)$, $\Delta f = 1/t_{msr}$, $t_{msr} = N \cdot \Delta t$ and $f_{Nyq} = 1/(2 \cdot \Delta t)$.

The imperfect fit in high frequency domain (~ 1.0 /min) indicates that the Lorentzian power spectrum is not a good estimator for the PS based on TPRW model.

Figures 2(d-f) exhibit the fitted motility parameters P, S and $\sigma_{pos}$ obtained from overall averaged MSD, VC and PS. Figure 2(d) indicates that all the fitted parameters P locate in the interval of 0.6 ~ 0.7 min, while the fitted parameters S in the interval of 0.2 ~ 0.25 μm/min. Neither of the resulting parameters P and S reflects the corresponding linear functional form of these parameters, they only approximately reproduce the corresponding averaged values ($P_{ave}$ = 0.65 min, $S_{ave}$ = 0.2 μm/min) of all theoretical P and S, respectively. The positioning errors $\sigma_{pos}$ in simulations are constant, thus we can average the fitted values from MSD and PS to obtain an approximate positioning error (~ 0.11 μm), instead of deriving the time-dependent errors.

### III. DERIVING THE ACCURATE TIME-VARYING MOTILITY PARAMETERS

In this section, we introduce wavelet transform and wavelet denoising to compute the wavelet power spectra of cell migration velocities, and further derive accurately time-dependent motility parameters via Lorentzian power spectrum.

#### A. WAVELET TRANSFORM OF MIGRATION VELOCITIES

The wavelet transform (WT) was initially employed by Morlet et al. to analyze seismic signals in the early 1980s [42,43], and was later formalized by Goupillaud and Grossmann et al. [44,45]. Different from the stationary process analyzed by Fourier transform (FT), the WT is regarded as a powerful tool to deal with the non-stationary and infinitely correlated process. For example, although the fractional Brownian motion is nonstationary and infinitely correlated, the corresponding wavelet coefficients are stationary and uncorrelated [46]. Kumar et al. also validated the incapability of FT to characterize the time-varying signals [47]. Further, windowed Fourier transform (WFT) can be computed by performing a sliding window of a constant time interval from a time series based on FT. It is also an analysis tool for extracting time-frequency information from a time series, but shows the inaccuracy and inefficiency because of the "imposed" window size into analysis (disadvantages, i.e., the most appropriate window size, the aliasing of high- and low-frequency, etc.), as discussed by Kaiser et al. [48], Torrence et al.[49] and Daubechies [50].

Different from the WFT, the window size can vary over the frequency in WT, which is the main advantage to analyze the local characteristics of time series [49-51]. Here, the WT includes discrete WT (DWT) and continuous WT (CWT), the latter is utilized in this study. For a given time series $v_n$, the CWT is defined as the convolution of $v_{n'}$ with a scaled and translated version of the wavelet function $\psi_0(\eta)$, as follows [49]

$$W_n(s) = \sum_{n'=0}^{N-1} v_{n'} \cdot \psi_0^* \left[ \frac{(n'-n) \cdot \Delta t}{s} \right], \tag{22}$$

where the symbol (*) denotes the complex conjugate, s is wavelet scale corresponding to Fourier frequency [49]. The wavelet function used here is Morlet, which consists of a plane wave modulated by a Gaussian

$$\psi_0(\eta)=\pi^{-1/4}\cdot e^{i\omega_0\eta}\cdot e^{-\eta^2/2}, \tag{23}$$

where $\omega_0$ is the non-dimensional frequency and is set as 6 for satisfying the admissibility condition [52]. Because the Morlet used above is complex, the resulting $W_n(s)$ is also complex. Thus, we can gain easily the information about the real part, imaginary part, and finally the wavelet power spectrum is computed by the absolute value squared of the wavelet transform [49,53], i.e., $|W_n(s)|^2$.

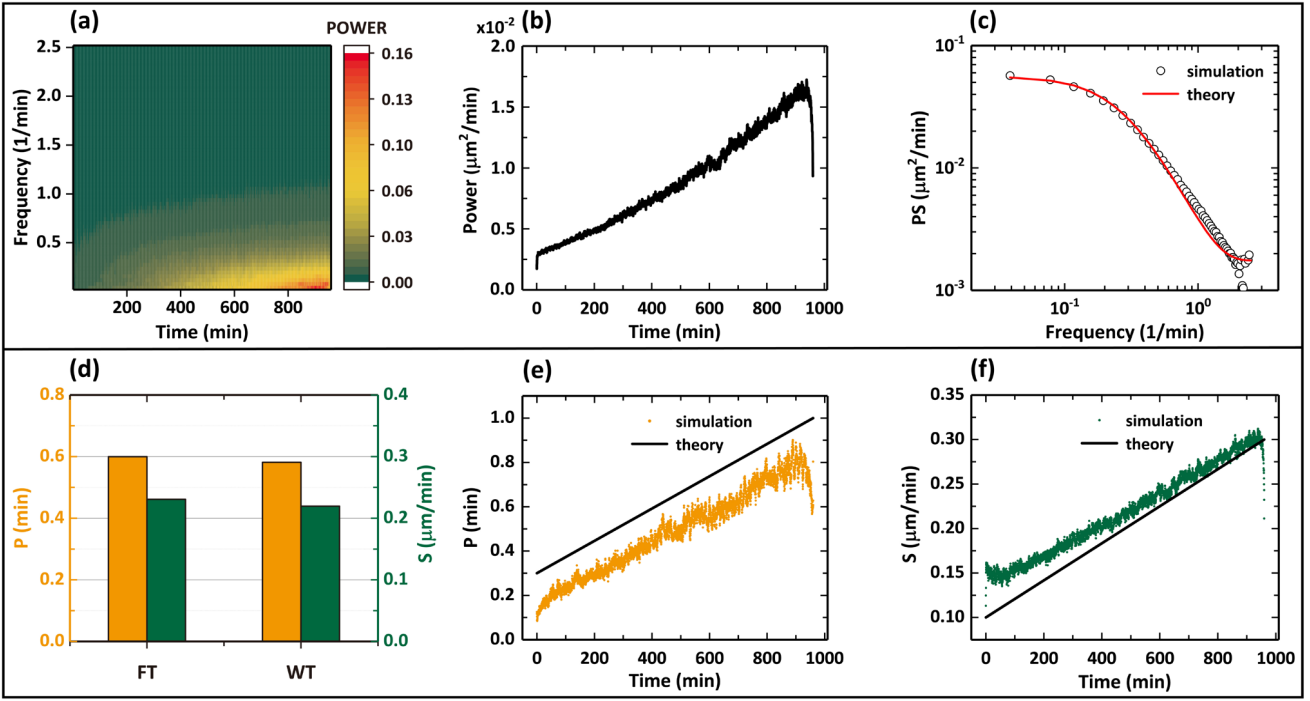

Fig. 3. Wavelet transform of cell migration velocities. (a) Wavelet power spectra based on Morlet wavelet function. The left axis denotes Fourier frequency corresponding to wavelet scale, while the bottom axis does time. The colors mirror the power spectral values. (b) The frequency-averaged wavelet power against time. (c) The time-averaged wavelet power spectrum (global wavelet power spectrum) against frequency. The red line represents theoretical values corresponding to Lorentzian power spectrum. (d) Comparison of the fitted parameters obtained from Fourier power spectrum and global wavelet power spectrum. The orange bars denote the persistence time P, while the green do the migration speed S. (e) The fitted persistence time P as a function of time. (f) The fitted migration speed S as a function of time. Note that the black lines stand for theoretical motility parameters in Fig. 1(a).

So far, one can follow the procedure above to graphically illustrating how the power spectral values change over the frequency and time, as illustrated in Fig. 3(a). In Fig. 3(a), the amplitude of wavelet power spectrum lies in the interval 0 ~ 0.16 μm²/min, which is denoted by different colors. At the same time, edge effects will occur in the beginning and end of the wavelet power spectrum because of the finite-length time series, which is also called as cone of influence (COI) [49], see the sharp decline parts affected by COI in Fig. 3(b). Note that we don't exclude the sharp decline parts for exhibiting the whole characteristics of the approach we developed.

**B. FITTING LOCAL WAVELET POWER SPECTRA WITH LORENTZIAN VELOCITY POWER SPECTRUM**

There is an evident peak with the time increasing in the low-frequency domain [see Fig. 3(a)], and it corresponds to the linear functions [see Fig. 1(a)]. Figure 3(a) shows not only the dominant features of migration velocities, but also how these features vary with time. The power spectra along frequency-axis for every moment in Fig. 3(a) are called as local wavelet power spectra, which are identical to the Fourier power spectra of the univariate lag-1 autoregressive [AR(1) or Markov] process, on average [49,54]. The 1D first-order AR(1) process is defined as

$$x_t = a_t \cdot x_{t-1} + b_t \cdot n_t, \tag{24}$$

where $a_t$ is the persistence parameter, $b_t$ the activity parameter and $n_t$ is obtained from Gaussian white noise. The process is equivalent to persistent random walk or an OU process. Here, the parameters $a_t$ and $b_t$ can be derived with Bayesian method proposed in Ref. [34], which is different from the motility parameters derived in this paper. When taking average on wavelet power spectra along frequency-axis, the frequency-averaged power (energy) is obtained, as shown in Fig. 3(b). The frequency-averaged power behaves like the peak in Fig. 3(a) and the tendency in Fig. 1(a), which means that the power of migrating cells increases with time, because of the enhancement of microenvironment. The abnormal decrease in the beginning and the end of the time series is a consequence of COI. When the average is taking over all the local wavelet power spectra, one will obtain the global wavelet power spectrum [see Fig. 3(c)], which is an unbiased and reliable estimation of the true power spectrum of any time series [55].

Further, Torrence et al. also validated the global wavelet power spectrum approximates to the corresponding Fourier power spectrum [49], thus it is reasonable to fit the global wavelet power spectrum using Lorentzian power spectrum mentioned in Sec. II, as shown by black line in Fig. 3(c). Figure 3(d) displays comparisons between motility parameters fitted from Fourier power spectrum and global wavelet power spectrum, respectively. It is obvious that the fitted parameters P and S based on WT are almost identical to these values based on FT, respectively. These identities further illustrate the rationality of Lorentzian power spectrum in fitting local wavelet power spectra.

There is no doubt that fitting the local wavelet power spectra using Lorentzian power spectrum will recover the time-dependent functions [see Figs. 3(e-f)]. The fitted parameters given in Fig. 3(e) for P and in Fig. 3(f) for S both encode the linear functions but with large deviations. Here, the fitted positioning errors are not shown, because they are not intrinsic terms related to cell motility due to the experimental observation. The corresponding errors can be estimated based on the results in Fig. 2(d).

**C. WAVELET DENOISING OF MIGRATION VELOCITIES**

In this part, we apply the wavelet denosing (WD) to filter the migration velocities before implementing the WT for improving the accuracy of fitting parameters. Here, WD mainly involves wavelet decomposition and reconstruction [56,57] and the main procedure is, first, we decompose the velocities to obtain the wavelet coefficients using Mallet algorithm [57]. Then, the coefficients are processed by a suitable threshold. Finally, the thresholded coefficients are reconstructed to obtain the denoised signal, as exhibited in Figs. 4(a) and 4(e). Here, the WD used is implemented based on Haar wavelet, which is the only discontinuous one of Daubechies wavelet family and known as the first order Daubechies wavelet db1, see more details in Ref. [58]. Moreover, the Haar wavelets are most commonly used wavelets in database literature because they are easy to comprehend and fast to compute [59,60], for instance, denoising the observed data without removing localized significant changes to represent the time-series evolution [61] and denoising for Haar wavelet to process the signal and image denoising [62]. When performing the wavelet decomposition once, the process is termed as "one layer" decomposition, while decomposition twice as "two layers". For simplicity, we use "1db1" to represent the process of denoising applying db1 wavelet with "one layer", while "2db1" to represent that with "two layers".

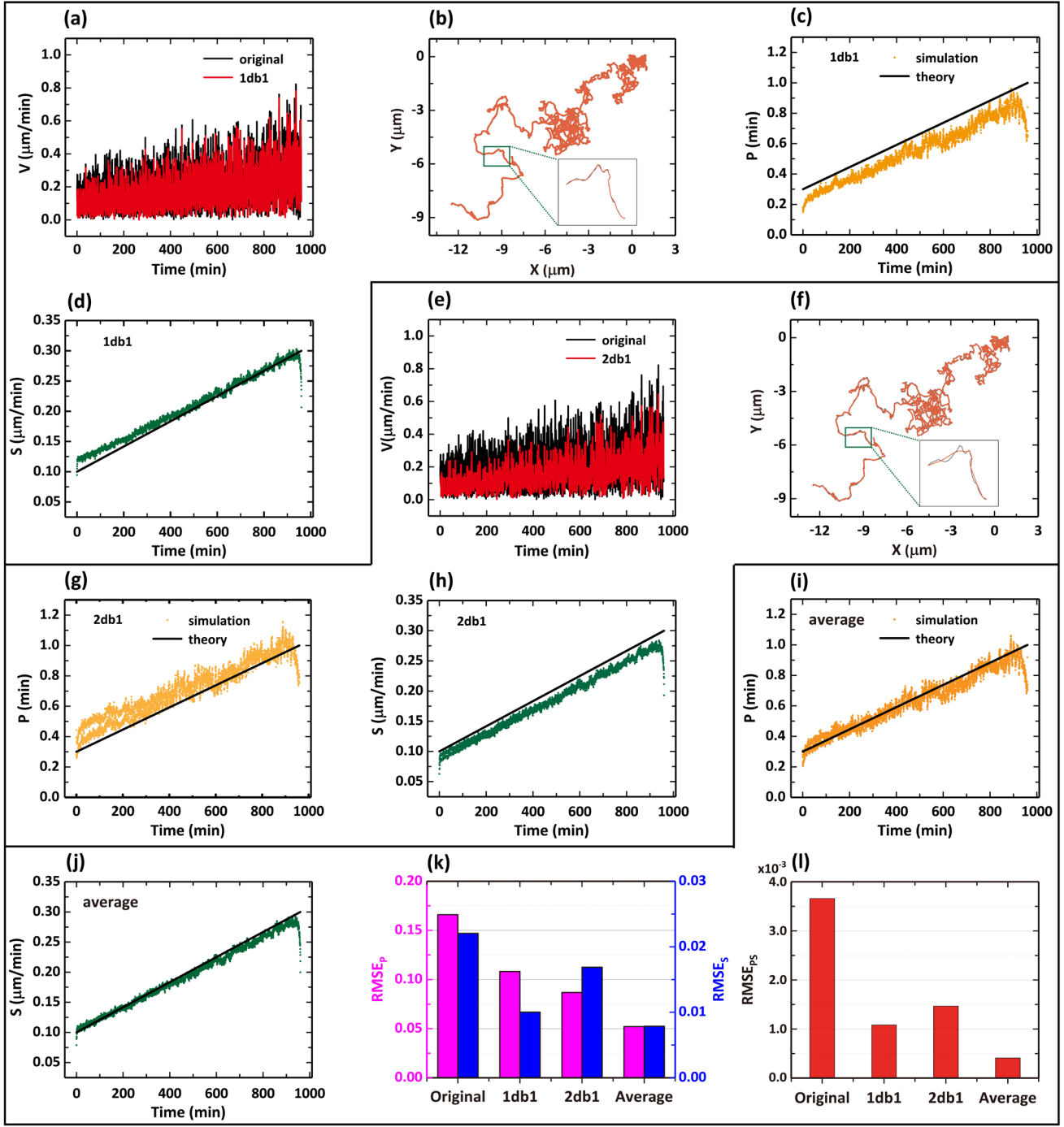

Fig. 4. Derivation of the accurate motility parameters based on wavelet denoising. (a) The migration velocities denoised by db1 wavelet function with one layer (1db1). The black line represents original migration velocities, while the red does the denoised. (b) The denoised migration trajectory corresponding to 1db1 and denoted by red line. The inset indicates the difference between the original and the denoised trajectories. The fitted persistence time P (c) and S (d) as functions of time in the case of 1db1. (e) - (f) The captions are identical to these for (a) - (b), but the results are obtained in the case of 2db1 (db1 wavelet function with two layers). (g) - (h) The captions are identical to these for (c) - (d), but in the case of 2db1 (db1 wavelet function with two layers). The persistence time P (i) and the migration speed S (j) obtained by averaging the corresponding results for the cases of 1db1 and 2db1. (k) Comparison of the RMSE ($RMSE_P$ and $RMSE_S$) of motility parameters corresponding to the four denoising methods (the original, 1db1, 2db1, the average). The mauve bars denote the $RMSE_P$ of the fitted persistence time

P, while the blue do the RMSE$_S$ of the fitted S. (l) Comparison of overall RMSE$_{PS}$. The overall RMSE$_{PS}$ are obtained by multiplying the RMSE$_P$ of P with RMSE$_S$ of S.

Figure 4(a) compares the original velocities and denoised velocities based on 1db1, which clearly shows the amplitude of the former is generally greater than that of the latter. Furthermore, the corresponding trajectories are exhibited in Fig. 4 (b). The enlarged inset illustrates the effect of 1db1 on trajectory, and the processed trajectory seems like more smoothly. Likewise, we employ 2db1 to denoise the same original velocities, and the results are seen in Figs. 4 (e) and 4(f). Comparing the denoised velocities in Figs. 4(a) and 4(e), it is evident that the 2db1 filters more velocity components (not just noise). Thus, the 2db1 smooth the trajectory more greatly, as shown by inset in Fig. 4(f). Note that cell migration trajectories discussed here contains typically two kind of noises, i.e., errors of observation and intrinsic part of their dynamics [25]. Actually, we do not know how much the noise component is for a migration trajectory.

For the migration velocities denoised by 1db1, we follow the same procedure used above, namely fitting local wavelet power spectra with Lorentzian power spectrum, to derive time-varying motility parameters P and S. The resulting parameters are plotted in Figs. 4(c) and 4(d), indicating P and S, respectively. The results for 1db1 are more accurate than these for original velocities in Figs. 3(e) and 3(f), which directly highlights the necessary of denoising. More importantly, we find that the fitted P are generally less than theoretical P, while the fitted S greater than theoretical S. we guess the "less and great" is the consequence of insufficient denoising, that is, the velocities denoised by 1db1 still contain some noise components, which decrease the persistence of cell migration. Next, we repeat the procedure but employing 2db1 to derive the motility parameters, as exhibited in Figs. 4(g) and 4(h). The results for 2db1 also are more close to theoretical values, but display opposite "great and less". Thus, we argue that the 2db1 is so "powerful" that more velocity components are filtered. This excessive denoising contributes to more persistent cell trajectories.

Inspired by the insufficient and excessive denoising, we further average these corresponding fitted parameters for 1db1 and 2db1, the averaged P and S are plotted in Figs. 4(i) and 4(j), respectively. The averaged results are exciting, because they are more accurate than these for the original, 1db1 and 2db1. What' more, the averaged results almost mirror the linear functions.

In order to compare the accuracy of the fitted parameters P and S based on different denoising methods, respectively, the corresponding root mean square errors (RMSE$_P$ and RMSE$_S$) of the fitted parameters P and S are calculated independently, as seen in Fig. 4(k). The bars indicate that both the 1db1 and 2db1 are better than the original, but the average is better than 1db1 and 2db1. To be more intuitive, overall RMSE$_{PS}$ are computed by multiplying RMSE (RMSE$_P$ and RMSE$_S$) of the fitted P and S [see Fig. 4 (l)], and it validates directly the advantage of the average in improving the accuracy of fitting motility parameters. Since the procedure used above mainly involves wavelet denoising, wavelet transform and Lorentzian power spectrum, we abbreviate it as WDTL for simplicity.

**D. TWO EXAMPLES OF MICROENVIRONMENT WITH TIME-VARYING CHARACTERISTICS**
**1. DEPENDENCY DEFINED BY QUADRATIC FUNCTIONS**

In the following, we discuss another two time-dependent functions to illustrate the universality of WDTL developed above. The functions for the first example are defined as

$$P(t) = K_P \cdot t^2 + P_0, \tag{25}$$

$$S(t) = K_S \cdot t^2 + S_0, \tag{26}$$

which are graphically shown in Fig. 5(a).

Next, the corresponding quantities for original migration velocities are computed and plotted in Figs. 5(b-f). They are comparable with Figs. 1 and 3. Further, we employ the WDTL to derive the time-dependent motility parameters, the averaged results are shown in Figs. 5(g) and 5(h). It is noticeable that the final values are consistent well with the theoretical values in Fig. 5(a). Figure 5(i) compares overall $RMSE_{PS}$ of the fitted parameters corresponding to the four denoising methods, which also illustrates that the average shows a higher performance in deriving the time-dependent motility parameters.

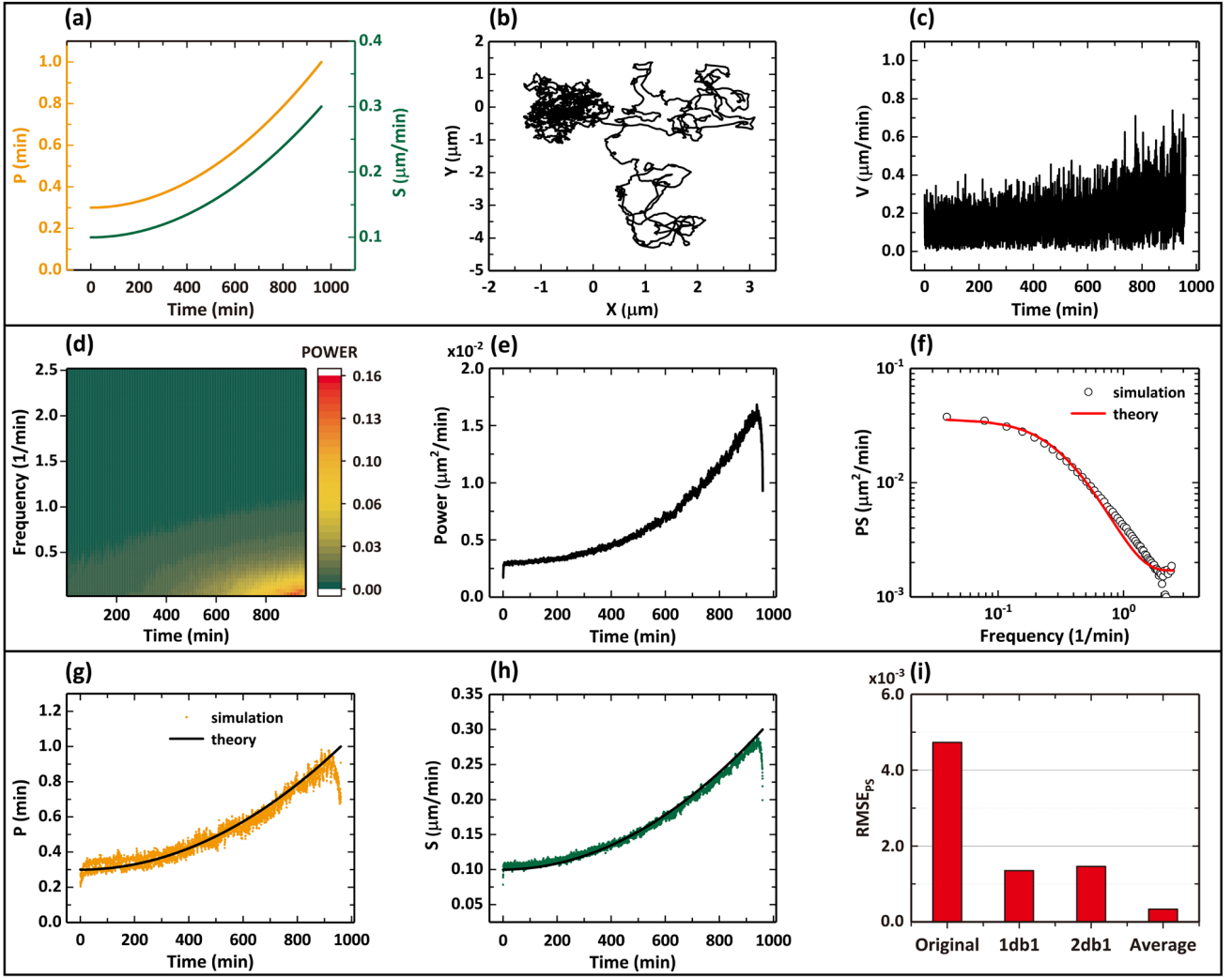

Fig. 5. Application of WDTL in deriving the accurate motility parameters following the quadratic functions. The captions are identical with these in Figs. 1, 3 and 4.

## 2. DEPENDENCY DEFINED BY COSINE FUNCTIONS

The time-dependent functions for the second example are given by

$$P(t)=A_P \cdot abs\left[\cos(2\pi \cdot t/T_P)\right]+P_0, \tag{27}$$

$$S(t)=A_S \cdot abs\left[\cos(2\pi \cdot t/T_P)\right]+S_0, \tag{28}$$

where $A_P$, $A_S$ are amplitudes, and $T_P$ is the period of the cosine. The detail values are shown in Fig. 6(a). Different from the linear and quadratic functions, the above functions not only contain the enhancement of microenvironment, but also the hindrance. Therefore, the cosine function provide a more realistic description of the real microenvironment with complex time-varying properties. The corresponding results are shown in Fig. 6. Figures. 6(g-i) also show the advantage of the average, which contributes to deriving the parameters reflecting the cosine dependencies.

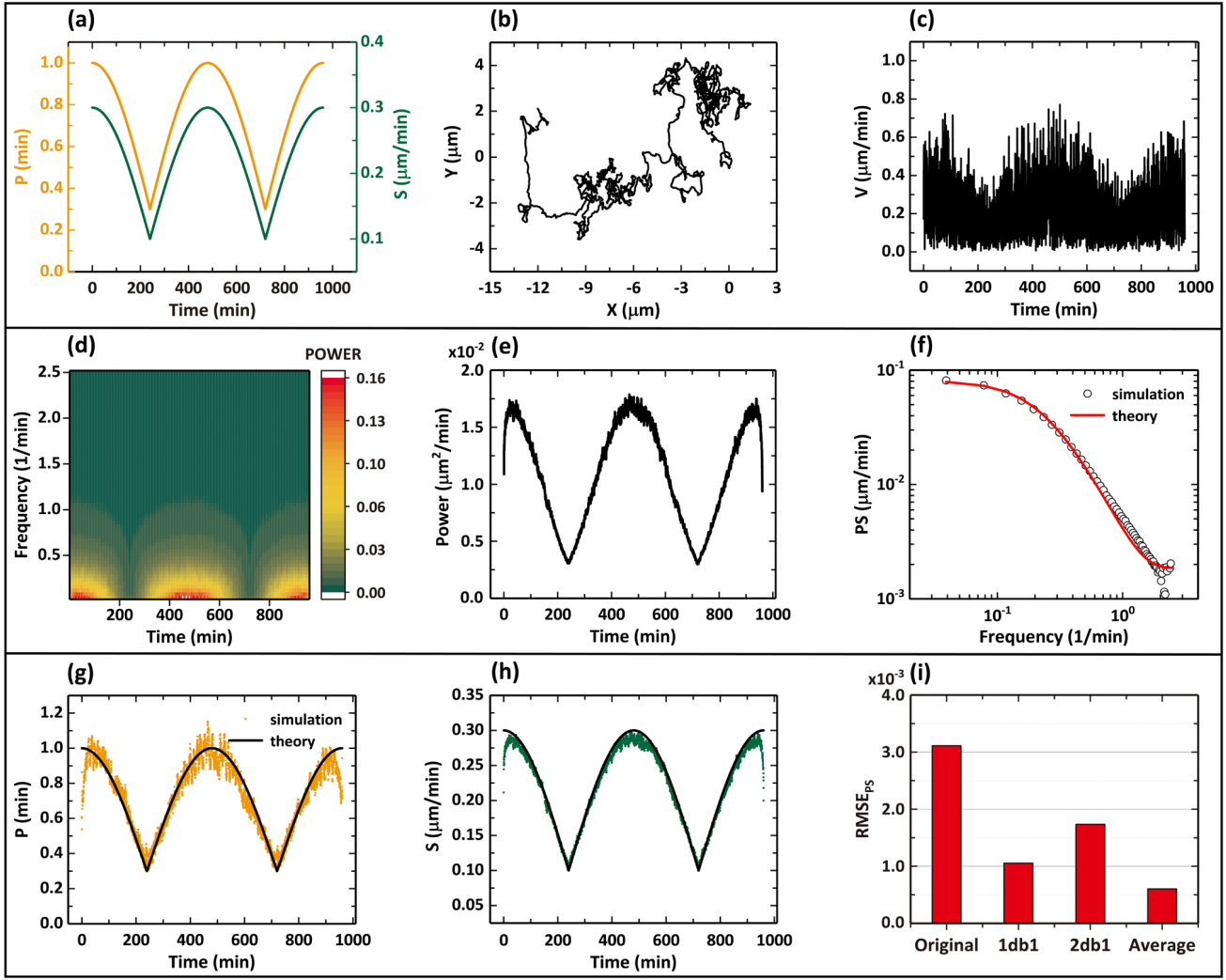

Fig. 6. Application of WDTL method in deriving the accurate motility parameters following the cosine functions. The captions are identical with these in Fig. 5.

Thus, we conclude that combination of wavelet denoising (1db1, 2db1 and the average), wavelet transform and Lorentzian power spectrum can derive more accurate motility parameters from migration velocities and further mirror the real-time changing microenvironment to some extent, which is also verified as a powerful analytical tool for cell motility.

**IV. THE EFFECTS OF SEVERAL FACTORS ON THE PERFORMANCE OF WDTL**

In the last section, we develop a robust and accurate approach (WDTL) to deriving the time-dependent motility parameters for different time-dependent functions. In order to further understand the limitations of WDTL, we continue to investigate the performance of WDTL under the influences of more factors based on Control Variates method and the linear functions, including the changing rates of microenvironmental properties (motility parameters) ($K_P$, $K_S$ and $K_{PS}$), the number of the recorded cells ($N_c$), the total recording time for individual trajectories (T), and the sampling time interval ($\Delta T$).

In the previous sections, we only study the performances of WDTL in the cases of three time-dependent functions, which are generally not sufficient to analyze the effects of various changing rates. Thus, we define several different changing rates [see Table 1], to study the effects of microenvironmental properties on the accuracy of fitting parameters. For simplicity, we again introduce linear functions [cf. Eqs. (2-3)], the corresponding quantities are computed and listed in Table 1. The indexes $K_P$ and $K_S$ quantify the changing rates of parameters P and S, respectively, while the $K_{PS}$ does the overall changing rates obtained from the average of the $K_P$ and $K_S$.

Table 1. The changing rates of microenvironmental properties (persistence time & migration speed)

| Quantities | Units | Group No. | | | | | | |
|---|---|---|---|---|---|---|---|---|
| | | 1 | 2 | 3 | 4 | 5 | 6 | 7 |
| $K_P$ | *(e-4) | 3.125 | 6.250 | 9.375 | 12.500 | 15.625 | 18.750 | 21.875 |
| $P_0$ | min | 0.3 | 0.3 | 0.3 | 0.3 | 0.3 | 0.3 | 0.3 |
| $P_T$ | min | 0.6 | 0.9 | 1.2 | 1.5 | 1.8 | 2.1 | 2.4 |
| $K_S$ | *(e-4) | 2.083 | 4.167 | 6.250 | 8.333 | 10.417 | 12.500 | 14.500 |
| $S_0$ | μm/min | 0.1 | 0.1 | 0.1 | 0.1 | 0.1 | 0.1 | 0.1 |
| $S_T$ | μm/min | 0.3 | 0.5 | 0.7 | 0.9 | 1.1 | 1.3 | 1.5 |
| $K_{PS}$ | *(e-4) | 2.604 | 5.209 | 7.813 | 10.417 | 13.021 | 15.625 | 18.188 |

Figures. 7(a-f) exhibit the detail results. When only increasing $K_P$ from 3.125e-4 to 21.875e-4 and keeping $K_S$ 2.083e-4, the $RMSE_P$ of fitted P increase and the fitted S almost does not be affected [see Fig. 7(a)]. The overall $RMSE_{PS}$ increase, as seen in Fig. 7(d). When increasing $K_S$ from 2.083e-4 to 14.5e-4 and keeping $K_P$ 7.292e-4, only the $RMSE_S$ of fitted S increase [see Fig. 7(b)]. The overall $RMSE_{PS}$ also increase [see Fig. 7(e)]. If we increase the $K_P$ and $K_S$ simultaneously, both the RMSE ($RMSE_P$ and $RMSE_S$) of P and S increase, as seen in Figs. 7(c) and 7(f).

Inversely, we find the overall $RMSE_{PS}$ decrease when increasing the number of the recorded cells [see Figs. 7(g) and 7(j)] or increasing the total recording time for individual trajectories [see Figs. 7(h) and 7(k)]. When keeping the total recording time T = 960 min constant but increasing the sampling time interval $\Delta T$ from 0.2 min to 2.0 min, the overall $RMSE_{PS}$ first decrease and then increase [see Figs. 7(i) and 7(l)].

We conclude that the slowly changing microenvironmental properties (~ 0), more recorded cells ( > 250), longer recording time ( > 600 min) and suitable sampling time interval (~ 0.6 min) will contribute to a better performance of our approach in fitting the time-dependent motility parameters.

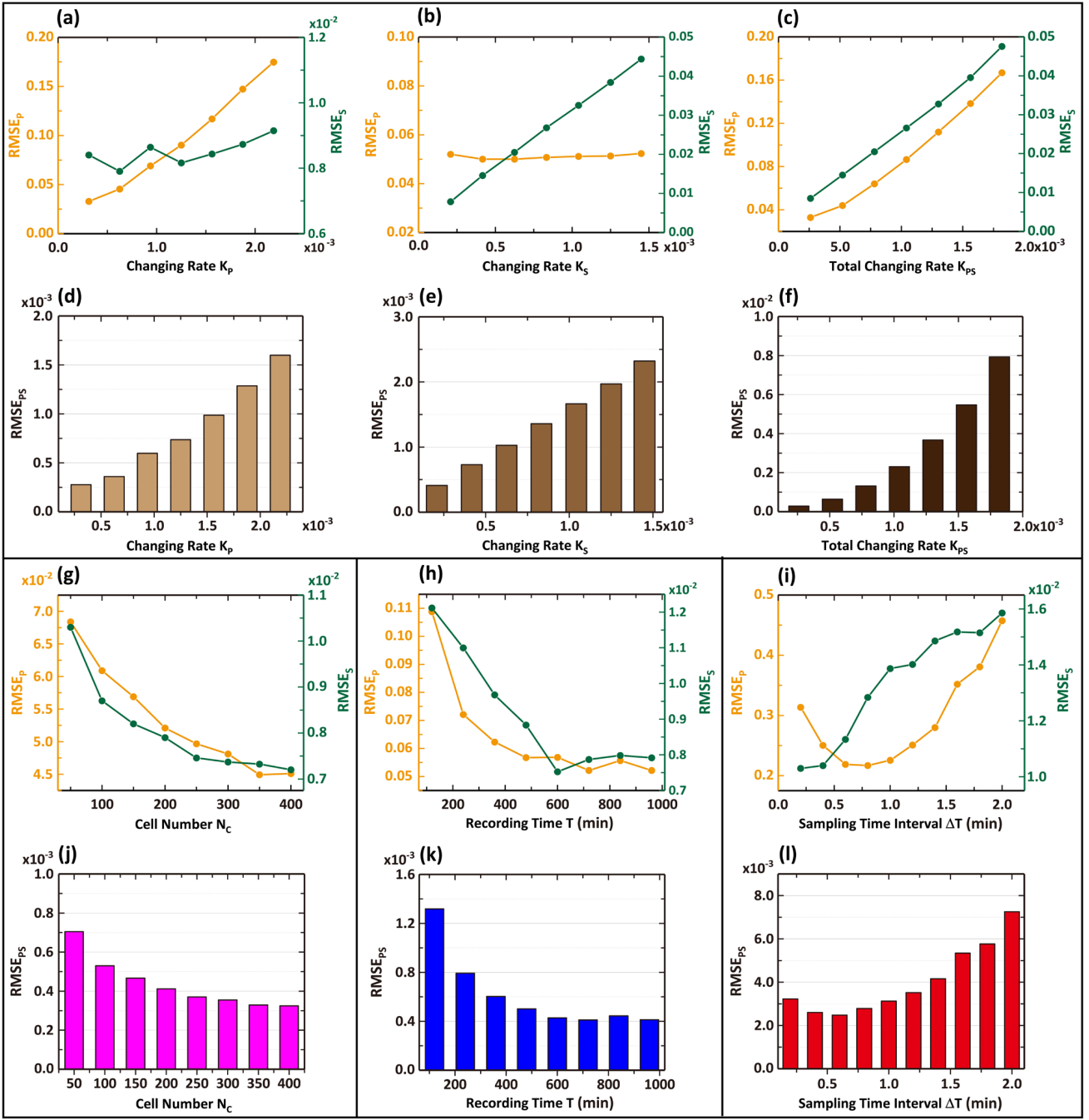

Fig. 7. The effects of several factors on the performance of WDTL method. (a) - (f) The effects of the changing rates of microenvironmental properties (motility parameters). (a) The RMSE of the fitted persistence time P and migration speed S as functions of the changing rates $K_P$. The orange denotes $RMSE_P$ of the fitted P, while the green does the $RMSE_S$ of the fitted S. (d) The overall $RMSE_{PS}$ as a function of the changing rates $K_P$. The captions of (b) and (e) are identical with these for (a) and (d), but for the changing rates $K_S$. (c) - (f) The $RMSE_P$ and $RMSE_S$ as functions of the overall changing rates $K_{PS}$ obtained by averaging the $K_P$ and $K_S$. (g) - (j) The RMSE as functions of the number of recorded cells $N_C$. (h) - (k) The RMSE as functions of the total recording time T for individual trajectories. (i) - (l) The RMSE as functions of the sampling time interval $\Delta T$.

## V. CONCLUSIONS AND DISCUSSION

Cell migration, which is of importance for the normal development of organisms and cancer metastasis, and is affected strictly by heterogeneous microenvironment. In this paper, we develop an approach (WDTL) to analyze the time-varying characteristics of cell migration, namely deriving the time-dependent motility parameters to reflect the changes of microenvironmental properties with time to some extent.

As a result, the cell motility parameters are the functions of time due to the influences of microenvironment, in the time-varying persistent random walk (TPRW). Based on trajectories simulated by TPRW model, we calculate MSD, VC and PS and further derive three sets of motility parameters (persistence time P, migration speed S and positioning error $\sigma_{pos}$) from the fits to the corresponding physical quantities. Although the three quantities can derive three sets of motility parameters, all of them only quantify the overall averaged cell migration capability instead of the time-dependent characteristics.

Then, we introduce the wavelet transform (WT) to compute the local wavelet power spectrum at each time step and obtain the time-dependent motility parameters by employing the Lorentzian power spectrum. However, the fitted results only roughly reflect the time-varying motility parameters with large deviations from the time-dependent functions. In order to improve the accuracy of WDTL, we apply wavelet denoising (WD) based on 1db1 and 2db1 to filter the migration velocities before implementing the wavelet transform. The results show clearly that the averaged parameters based on 1db1 and 2db1 can significantly decrease the errors between the fitted and the theoretical motility parameters, mirroring the time-dependent functions. In order to verify this approach, we further analyze the microenvironment with quadratic and cosine functions, the results show that the approach still exhibits higher performances in fitting motility parameters.

Therefore, we conclude that the combination of wavelet denoising, wavelet transform and Lorentzian power spectrum can derive nicely the motility parameters from cell trajectories to sufficiently mirror the time-dependent characteristics of cell migration which can reflect the varying properties of microenvironment with time, to some extent.

In addition, we continue to study the effects of several factors on the performances of WDTL, for instance, the changing rates of microenvironmental properties (motility parameters), the number of the recorded cells, the total recording time for individual trajectories and the sampling time interval. The studies figure out the limitations of the WDTL developed, which also provide a guidance when processing the cell migration in Lab.

We should first point out that the wavelet function is very critical to in the process of denoising, but we only choose the commonly used db1 instead of others, which may not be the best choice. Second, when the overall average of 200 cells is performed, the individual differences of the cells are averaged out. Finally, although one can derive the time-dependent motility parameters with WDTL, but how these varying parameters actually reflect the changes of microenvironment characteristics remains unclear, which is a future research interest.

**ACKNOWLEDGMENTS**

This research was supported by the National Natural Science Foundation of China (Grant Nos. 11974066, 11674043, 11675134, 11874310), the Fundamental Research Funds for the Central Universities (Grant No. 2019CDYGYB007), and the Natural Science Foundation of Chongqing, China (Grant No. cstc2019jcyj-msxmX0477, cstc2018jcyjA3679). Y. J. thanks the support from Arizona State University and hospitality from Peking University during his sabbatical leave.